\documentclass{elsart}
\usepackage{epsfig}
\usepackage{amssymb}

\def\cal{\mathcal}
\def\piz{\pi ^0 }
\def\pip{\pi ^+ }

\def\BBbar{B\bar{B}}
\def\epem{e^+ e^-}

\def\Emiss{E_{\rm miss}}
\def\pmiss{\vec{p}_{\rm miss}}

\def\GeV{{\rm GeV}}
\def\GeVc{{{\rm GeV}/c}}
\def\GeVcc{{{\rm GeV}/c^2}}
\def\MeV{{\rm MeV}}

\def\MeVcc{{\rm MeV/}c^2}
\def\to{\rightarrow}
\def\BR{{\cal B}}

\def\Vcb{V_{cb}}

\def\MB{M_{\rm bc}}
\def\BDlnu{\bar{B^0} \rt D^{+}\ell^-\bar{\nu}}
\def\BDpzerolnu{\bar{B} \rt D\ell^-\bar{\nu}}

\def\BDstlnu{\bar{B^0} \rt D^{*+}\ell^-\bar{\nu}}
\def\BDstlnuall{\bar{B} \rt D^{*}\ell^-\bar{\nu}}
\def\y{y}

\def\mybr{2.13\pm0.12\pm0.39}
\def\mygamma{13.79\pm0.76\pm2.51}

\def\myvcbfcap{4.11\pm0.44\pm0.52 }
\def\myvcb{4.19\pm0.45\pm0.53\pm0.30 }

\newcommand{\rt}{\rightarrow}
\newcommand{\vcb}{|V_{cb}|}

\newcommand{\vcbf}{|V_{cb}|F_{D}(1)}
\newcommand{\vcbfw}{|V_{cb}|F_{D}(y)}
\newcommand{\rsqr}{\hat{\rho}^2_{D}}

\begin{document}
\begin{frontmatter}

\begin{flushleft}\includegraphics[width=3.5cm]{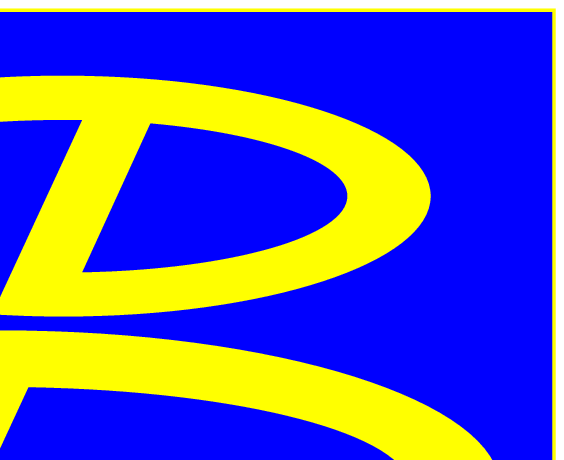}\end{flushleft}
\vspace{-2.5cm}
\vbox{\normalsize%
\noindent%
\rightline{\hfill {\tt KEK preprint 2001-150}}\\%
\rightline{\hfill {\tt Belle preprint 2001-20}}%
}
\vspace{2.5cm}





\title{
Measurement of ${\cal B} (\BDlnu)$ and Determination of $\vcb$}

\author{Belle Collaboration}

\begin{center}
{\normalsize
  K.~Abe$^{9}$,               
  K.~Abe$^{43}$,              
  R.~Abe$^{32}$,              
  T.~Abe$^{44}$,              
  I.~Adachi$^{9}$,            
  Byoung~Sup~Ahn$^{17}$,      
  H.~Aihara$^{45}$,           
  M.~Akatsu$^{25}$,           
  Y.~Asano$^{50}$,            
  T.~Aso$^{49}$
% ...
}\end{center}

\address{
% ...
$^{9}${High Energy Accelerator Research Organization (KEK), Tsukuba}\\
% ...
$^{17}${Korea University, Seoul}\\
% ...
$^{25}${Nagoya University, Nagoya}\\
% ...
$^{32}${Niigata University, Niigata}\\
% ...
$^{43}${Tohoku Gakuin University, Tagajo}\\
$^{44}${Tohoku University, Sendai}\\
$^{45}${University of Tokyo, Tokyo}\\
% ...
$^{49}${Toyama National College of Maritime Technology, Toyama}\\
$^{50}${University of Tsukuba, Tsukuba}\\
% ...
}

\normalsize
\begin{abstract}

We present a measurement of the branching fraction for the semileptonic $B$ decay $\BDlnu$, where $\ell^-$ can be either an electron or a muon. We find  $\Gamma(\BDlnu)=(\mygamma)~{\rm ns^{-1}}$, and the resulting branching fraction  $\BR (\BDlnu)= (\mybr)\%$,
 where the first error is statistical and the second systematic.
We also investigate the $\BDlnu$ form factor and the implications of the result for $\vcb$. From a fit to the differential decay distribution we obtain 
the rate normalization $\vcbf=(\myvcbfcap)\times10^{-2}$. 
Using a theoretical calculation of
$F_D(1)$, the Cabibbo-Kobayashi-Maskawa matrix element  $|\Vcb|=(\myvcb)\times10^{-2} $ is
obtained, where the last error comes from the theoretical uncertainty of
$F_D(1)$.
The results are based on a data sample of 10.2$~{\rm fb^{-1}}$ recorded at the $\Upsilon(4S)$ resonance with the Belle detector at the KEKB $e^+e^-$ collider. 
\end{abstract}


\end{frontmatter}

\clearpage

\section{Introduction}

In the Standard Model of electroweak interactions, the elements
of the Cabibbo-Kobayashi-Maskawa (CKM) quark mixing matrix~\cite{CKM}
are constrained by unitarity.  Therefore, experimental measurements of
the precise values of the CKM matrix elements are important to
understand the phenomenology of weak interactions.

In the framework of Heavy Quark Effective Theory (HQET)~\cite{HQET},
the semileptonic decay $\BDlnu$ is amongst the cleanest modes that can be used
to measure the CKM matrix element $|\Vcb|$. The differential decay rate for $\BDlnu$ can be expressed as~\cite{Neubert}

\begin{equation}
 \frac{d\Gamma}{d\y} = \frac{G_F^2 |V_{cb}|^2}{48\pi^3}(m_{\bar{B}^0}+ m_{D^+})^2 m_{D^+}^3 (\y^2-1)^{3/2} F_D^2(\y), 
\end{equation}
where $G_F$ is the Fermi coupling constant, and $m_{\bar{B}^0}$ and $m_{D^+}$ are the masses of the $\bar{B}^0$ and $D^+$ mesons respectively. The variable $\y$ denotes the inner product of the $\bar{B}^0$ and $D^+$ meson four-velocities, which is related to $q^2$, the mass squared of the lepton-neutrino system:
\begin{equation}
 \y =  v_{\bar{B^0}} \cdot v_{D^+}= \frac{m_{\bar{B}^0}^2 + m_{D^+}^2 - q^2}{2m_{\bar{B}^0} m_{D^+}},~ q^2 = (p_{l^-} + p_{\bar{\nu}})^2,  
\label{omega}
\end{equation}
 and $F_D(\y)$ is the form factor. By extrapolating the measured $\vcbfw$ to the point of zero recoil of the $D^+$ meson, $\vcb$ can be determined using a theoretical prediction of $F_D(1)$.  In this paper $p$ represents the four-momentum vector of the particle in subscript, while $\vec{p}$ denotes the three-momentum vector in the center of mass (CM) frame and $\vec{p}_{lab}$ is the three-momentum vector in the laboratory frame.

$|\Vcb|$ measurements made using $\BDpzerolnu$ decays are less
precise than those made using $\BDstlnuall$, due to the suppressed decay rate near the point of zero recoil, substantial feed down from $\BDstlnuall$ and the large combinatoric background from fake $D$ mesons. 
Nevertheless, the $\BDpzerolnu$ measurement provides a consistency check, and allows a test of heavy-quark symmetry~\cite{hqsymmetry} through the precise measurement of the form factor. Furthermore, this  decay has one experimental advantage over $\BDstlnuall$ as there is no slow pion involved.

 The $\BDlnu$ decay is preferred to $B^-\to D^0 \ell^-\bar{\nu}$ 
because it has much less feed-down background from excited
charm meson states. 
Good particle identification and vertexing capabilities allow 
effective reduction of the combinatoric background.

In this paper we report measurements of the branching fraction
of the semileptonic decay $\BDlnu$, the CKM matrix element $|V_{cb}|$ and the form factor $F_D(\y)$.  The lepton $\ell$ can be either an electron or a muon, and the
use of the charge-conjugate mode is implied.

The data sample used in this analysis was
collected with the Belle detector~\cite{NIM} at KEKB~\cite{kekb}, an asymmetric $e^+ e^-$ collider.
The data sample has an integrated luminosity of 
$10.2$ fb$^{-1}$ and contains $10.8 \times 10^{6}$ 
$B \bar{B}$ pairs. 
Another data sample with an integrated luminosity of $0.6 $ fb$^{-1}$
was taken at an energy 60 MeV below the $\Upsilon(4S)$ resonance, 
and is used as a control sample to check the $\epem \to q\bar{q}$ continuum background determination.

\section{Belle detector}

Belle is 
a large solid-angle spectrometer based on a 1.5~T superconducting
solenoid magnet.  Charged particle tracking is provided by a
silicon vertex detector (SVD) and a central drift chamber
(CDC) that surround the interaction region.   The SVD consists
of three approximately cylindrical
layers of double-sided silicon strip detectors;
one side of each detector measures the $z$ coordinate and the
other the $r-\phi$ coordinate.
The CDC has  50 cylindrical layers of anode
wires; the inner three layers have instrumented cathodes
for $z$ coordinate measurements.
Eighteen of the wire layers are inclined at small angles 
to provide small-angle stereo
measurements of $z$ coordinates along the particle trajectories.
The charged particle acceptance covers the laboratory polar angle between
$\theta=17^\circ$ and $150^\circ$, corresponding to about 92\%
of the full CM frame solid angle.
The momentum resolution for charged tracks
is determined from cosmic rays and $e^+ e^-\to\mu^+\mu^-$ events
to be $(\sigma_{|\vec{p}_t|}/ |\vec{p}_t|)^2 = (0.0019{|\vec{p}_t}|)^2 + (0.0030)^2$, where $\vec{p}_t$ is the transverse momentum in $\GeVc$. 

Charged hadron identification is provided by $dE/dx$ measurements in
the CDC,  a mosaic of 1188 aerogel \v{C}erenkov counters
(ACC), and a barrel-like array of 128
time-of-flight scintillation counters (TOF).  The $dE/dx$ 
measurements have a resolution
for hadron tracks of 6.9\% and are useful for $\pi /K$ separation
for $|\vec{p}_{lab}| < 0.8~\GeVc$ and $|\vec{p}_{lab}| > 2.5~\GeVc$.
The TOF system
has a time resolution for hadrons of $\sigma \simeq 100$~ps and
provides $\pi /K$ separation for  $|\vec{p}_{lab}| < 1.2~\GeVc$. 
The ACC covers the range $1.2~\GeVc < |\vec{p}_{lab}| < 3.5~\GeVc$, and the refractive indices of the ACC elements vary with
 polar angle to match the kinematics of the
asymmetric energy collisions.
High momentum tagged kaons and pions from
kinematically selected $D^{*+}\to D^0\pi^+$, $D^0\to K^-\pi^+$
decays are used to determine a charged kaon identification
efficiency of 88\% and a misidentification probability of 9\%.

Electromagnetic showering particles are detected in
an array of 8736 CsI(Tl) crystals located in the magnetic
volume covering the same solid angle as the charged particle
tracking system.  The energy resolution for
electromagnetic showers is 
$(\sigma_E/E)^2 = (0.013)^2 + (0.0007/E)^2 + (0.008/E^{1/4})^2$, where $E$ is in GeV. 
Neutral pions are detected via their $\pi^0\to\gamma\gamma$ decay. 
The $\pi^0$ mass resolution varies
slowly with energy, averaging $\sigma = 4.9~\MeVcc$.
For a $\pm 3\sigma$ mass selection requirement, the overall
detection efficiency for
neutral pions from $B\bar{B}$ events 
(including the effects of geometrical acceptance)
is $\sim 40\%$.

Electron identification in Belle is based on
a combination of $dE/dx$ measurements
in the CDC, the response of the ACC,
the position and shape of the
electromagnetic shower, and the ratio $E/|\vec{p}_{lab}|$ of cluster energy registered in the calorimeter and particle momentum.  The electron identification efficiency, determined by embedding Monte Carlo (MC)~\cite{MC}  tracks in multihadron data, is greater than 92\% for tracks with $|\vec{p}_{lab}|>1.0~\GeVc$. The hadron misidentification probability, determined from $K_S^0\to \pi^+\pi^-$ decays, is below 0.3\%.

The 1.5~T magnetic field is returned via an iron
yoke that is instrumented to detect 
muons and $K_L^0$ mesons.  This detection system 
consists of alternating layers of charged-particle detectors and
4.7~cm thick steel plates.  The total steel thickness 
of 65.8 cm
plus the material of the inner detector corresponds to 4.7 nuclear
interaction lengths at normal incidence. The system covers 
polar angles between $\theta=20^\circ$ and $155^\circ$; the
overall muon identification efficiency, determined by a track
embedding study similar to that used for the electron case,
 is greater than 87\% for tracks with $|\vec{p}_{lab}| > 1~\GeVc$.
 The corresponding pion misidentification probability
determined from inclusive $K_S^0 \to \pi^+\pi^-$ decays is less than 2\%.

\section{Event selection and analysis procedure }

\subsection{Event selection}
For hadronic event selection, we require that each event
have at least 5 well-reconstructed charged tracks, a total
visible energy of at least 0.15 times the CM energy
and an event vertex that is consistent with the known interaction
point.  
Continuum background events are suppressed by requiring
the normalized second Fox-Wolfram moment~\cite{FoxWolf} to be less
than 0.4.

This analysis  is based on the neutrino reconstruction method, which exploits the hermeticity of the detector and the near zero value of the neutrino mass.  This method was originally developed by the CLEO collaboration for the measurement of $B\to \pi \ell\nu$ and $\rho (\omega) \ell\nu$ decays~\cite{CLEOPilnu}.

We extract information on the neutrino from the missing momentum ($\pmiss$)  and missing energy ($\Emiss$) in each event. In the CM frame, the total momentum of the system is zero, and the total energy is the sum of the two beam energies ($E_{\rm beam}$).  The
missing energy, momentum and missing mass are calculated as follows:
\begin{equation}
 \Emiss  = 2E_{\rm  beam}  - \Sigma  E_{i},
\end{equation}
\begin{equation}
 \pmiss = - \Sigma  \vec{p}_{i},
\end{equation}
\begin{equation}
 M_{\rm  miss}^2 =  E_{\rm  miss}^2 - |\pmiss| ^2,
\end{equation}

where the sums are over all reconstructed particles $i$ in the event. 

In calculating the missing energy and momentum, we identify each charged particle using the lepton and hadron identification devices.
We also add the neutral showers recorded in the ECL that are not matched to
any charged track.  Moreover, we require that the shower shape be
consistent with that of a photon and the deposited energy be greater
than $30~\MeV$ for the barrel region and $50~\MeV$ for the endcap.

If the only undetected particle in an event is the neutrino,
the missing mass should be consistent with zero. Due to the presence of a tail in the missing mass distribution originating from missing particles, an asymmetric requirement $-2.0~{{\rm GeV^2}/c^4} <M_{\rm miss}^2<3.0~{{\rm GeV^2}/c^4}$ is applied. 
In principle, this selects events where 
there is only one undetected particle in the event and 
that missing particle is the neutrino.  
Neutrinos are usually produced in conjunction with charged leptons ($e$ or $\mu$). 
Therefore, the presence of two or more leptons in an event implies that
the missing energy and momentum cannot be used to accurately reconstruct
the neutrino from the $\BDlnu$ decay. 
For this reason,
we select events with only one identified lepton  with
$|\vec{p}_{lab}|>0.8~\GeVc$.  
The momentum  requirement helps to
reduce the signal 
efficiency loss due to lepton misidentification, and also
reduces the backgrounds from hadrons misidentified as leptons and random
combinations of leptons and $D^+$ mesons. 

Since the initial state is charge-neutral, the charges of the reconstructed tracks should sum to zero if we detect all the particles in the event and have no additional tracks.
Hence, we require that the net charge,
$\Delta Q$, in the event should be close to zero to reject events
with other missing charged particles.  We allow $\Delta Q=\pm 1$, as well as 0, to maintain  high signal efficiency.

The neutrino reconstruction can be biased by particles
that go undetected by passing down the beam pipe.  To reject events with
such missing particles, we require 
$\left| \cos \theta_{\vec{p}_{\rm miss}} \right| <0.95$, where $\theta_{\vec{p}_{\rm miss}}$ is the
polar angle of the missing momentum with respect to the positron beam
direction in the laboratory frame.

\begin{figure}[ht] 
\centerline{
\epsfysize=4.5in 
\epsfbox{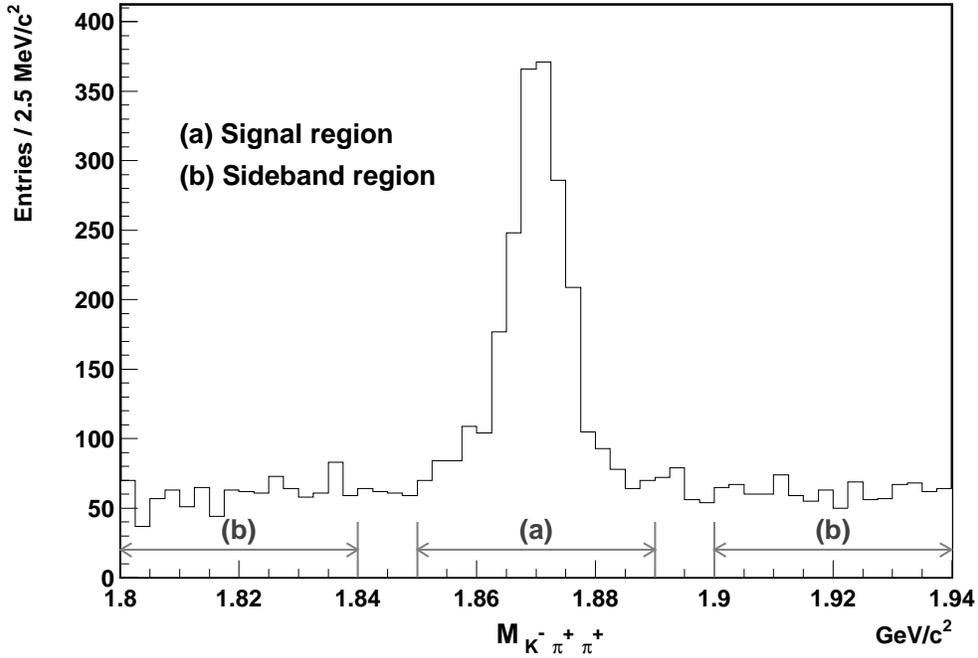}
}
\caption{\leftskip=0.25truein \rightskip=0.25truein $M_{K^- \pi^+ \pi^+}$
distribution after all event selection criteria except the $M_{K^- \pi^+ \pi^+}$ requirement. The signal region is indicated by (a) and the sideband regions by (b).}
\label{mdplot}
\end{figure}

$D^+$ candidates are reconstructed in the $D^+ 
\rightarrow K^- \pi^+ \pi^+$ decay mode.  
Kaon candidates are required to be positively identified by the hadron 
identification devices. In addition, we require kaon candidates not to be
positively identified as either a lepton or a proton.  If a charged
particle is not positively identified as a lepton, kaon, or proton,
we treat it as a pion.  The three charged tracks are then
geometrically fit to a $D^+$ decay vertex, and we reject combinations which do not form a consistent vertex.
We further require $|\vec{p}_{D^+}|<2.5~\GeVc$ in order to suppress continuum background. 
We select $K^-\pi^+\pi^+$ combinations
where the invariant mass is within 20 ${\MeVcc}$ of the nominal
$D^+$ mass (Figure~\ref{mdplot}).

To reduce the feed-down background from $\bar{B} \to D^{*+} X \ell^-\bar{\nu}$ decays,
we reject $D^+$ candidates that are consistent with being produced in
the decay $D^{*+}\to D^+\pi^0$.  We select $\piz$ candidates by requiring that
the two photon invariant mass be within $16.5~\MeVcc$ of the nominal
$\piz$ mass, where the energy of each photon is required to be greater than
20~$\MeV$.  We calculate the mass difference
$M_{K^-\pip\pip\piz}-M_{K^-\pip\pip}$ and if it is within $2~\MeVcc$
of the nominal value for a $D^{*+}$ decay, the $D^+$ candidate is
rejected.

A variable $\cos \theta_{B-D\ell}$ is defined as the cosine of the
angle between $\vec{p}_{\bar{B}^0} $ and $\vec{p}_{D^+ \ell^-} (=
\vec{p}_{D^+}+\vec{p}_{\ell^-})$. It satisfies the following
kinematic relation:
\begin{equation}
  \cos \theta_{B-D\ell} = \frac{2 E_{\bar{B}^0} E_{D^+\ell^-} - m^2_{\bar{B}^0 }- M^2_{D^+\ell^-}}{2|\vec{p}_{\bar{B}^0}||\vec{p}_{D^+\ell^-}|}.
\end{equation}
The signal events are distributed mostly within the physically allowed region $\left| \cos \theta_{B-D\ell} \right| <1 $, while
the background events extend to a much wider range.  We require that
candidates have $\left| \cos \theta_{B-D\ell} \right| < 1$.

Since the resolution of the missing momentum is
 better than that of the missing energy, we take 
$(E_{\bar{\nu}},\vec{p}_{\bar{\nu}}) = (|\vec{p}_{\rm
miss}|,\vec{p}_{\rm miss})$ as the four-momentum of the neutrino.
Combining the energy-momentum four-vectors for the reconstructed $D^+$
meson, the signal lepton and the neutrino, and using the constraint
of energy-momentum conservation, we obtain the fully reconstructed $B$
decay variables, the beam constrained mass $\MB$
and the energy difference $\Delta E$ defined as
\begin{equation}
 \Delta E = (E_{D^+}+E_{\ell^-}+E_{\bar{\nu}})-E_{\rm beam},
\end{equation}
\begin{equation}
 \MB = \sqrt{ E^2_{\rm beam} -
 |\vec{p}_{D^+}+\vec{p}_{\ell^-}+\alpha\vec{p}_{\bar{\nu}}|^2 }.
 \label{EqMB} 
\end{equation}

We select events with $\Delta E$ in the range $-0.2~\GeV
<\Delta E < 1~ \GeV $; the asymmetric requirement is to reject feed-down background from $\bar{B} \to D^{*+} X \ell^-\bar{\nu}$ decays. In the calculation of $\MB$ (Equation~\ref{EqMB}) we correct $p_{\bar{\nu}}$ by the factor
\begin{equation}
  \alpha = 1 + \Delta E/E_{\bar{\nu}},
\label{alpha}
\end{equation}
which is equivalent to imposing $\Delta E = 0$.

\subsection{Background sources}
The background sources fall into five categories: combinatoric,
correlated, uncorrelated, misidentified lepton and continuum. 

{\it Combinatoric background\rm:} The dominant background in this analysis
is the combinatoric background in the $D^{+}$ reconstruction.  
The amount of combinatoric background is
estimated using the events in the sideband regions $1.80~\GeVcc <M_{K^- \pip\pip} <1.84~\GeVcc$ and $1.90~\GeVcc <M_{K^- \pip\pip} <1.94~\GeVcc$, shown in Figure~\ref{mdplot}. 

{\it Correlated background\rm:} If a $D^{+}$ and a lepton have the same
parent $B$, but do not come from the signal decay $\BDlnu$, the event is classified as  correlated background.  Processes such as  $\BDstlnu$, ${\bar B} \rightarrow D^{**}\ell^- {\bar \nu}$ and ${\bar B} \rightarrow D^{(*)}\pi\ell^- {\bar \nu}$ (nonresonant) contribute to this source. This background is estimated by MC simulation. 
We use the measured form factor~\cite{DstarFF} and decay rate~\cite{pdg} for $\BDstlnu$ which constitutes the majority (89\%) of this background. For ${\bar B} \rightarrow D^{**}\ell^- {\bar \nu}$ and ${\bar B} \rightarrow D^{(*)}\pi\ell^-{\bar\nu}$, the models in refs.~\cite{ISGW2} and~\cite{goity} are assumed, respectively.
Our assumptions for these decay rates~\cite{BrAssumption} are based on existing measurements~\cite{pdg}.

{\it Uncorrelated background\rm:} The uncorrelated background consists of events with
a real $D^{+}$ from the decay of one $B$ meson and a real lepton
from the opposite $B$. We also estimate this background using a MC simulation. Due to the charge correlation with the $D^+$, the
lepton in this background is usually from a secondary decay, and is somewhat suppressed by the lepton momentum requirement. 

{\it Misidentified lepton background\rm:} In some cases, a hadron track is
misidentified as a lepton.  The amount of this background is estimated from data.
We treat each hadron candidate track as if it were a signal lepton and
weight its contribution according to the misidentification probability
measured using kinematically identified hadron tracks in 
data. 

{\it Continuum background\rm:} The $\epem \to q\bar{q}$ continuum
background is estimated using MC continuum events.  We also
check that this estimation is statistically consistent with
the off-resonance data.

\subsection{Signal yield and efficiency}

\begin{figure}[ht] 
\centerline{
\epsfysize=4.5in 
\epsfbox{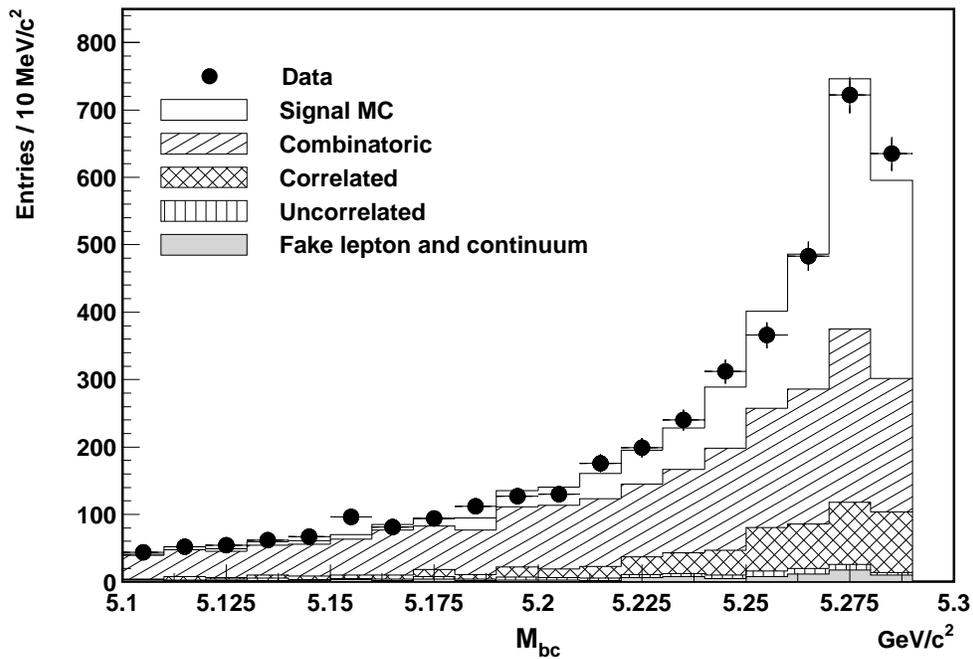}
}
\caption{\leftskip=0.25truein \rightskip=0.25truein $\MB$
distribution and estimated backgrounds. 
The points with error bars show the data. The shaded component is the continuum and fake lepton background, the vertically hatched histogram is the uncorrelated background, the cross-hatched histogram is the correlated background, and the diagonal hatched histogram is the combinatoric background. The open histogram is the signal MC, normalized to the measured decay rate.}
\label{mbplot}
\end{figure}

Figure~\ref{mbplot} shows the
$\MB$ distribution after all the event selection criteria described above. In this plot the points with error bars represent the on-resonance data. The background components are also shown.

We apply a final event selection requirement $\MB > 5.24~\GeVcc$ to define the signal region.
The overall signal efficiency is $2.69\%$. Table~\ref{table:signal} lists the number of signal events and the estimated backgrounds in the signal region.

\section{Measurement of \boldmath{${\vcbf}$}} 

The uncertainty on the reconstructed value of $\y$ is dominated by the error on the measurement of the neutrino four-momentum.  We denote 
the reconstructed value as $\tilde{\y}$. The resolution is improved by making the correction to the momentum measurement by the factor $\alpha$ given in Equation \ref{alpha}.
Using a MC simulation, the resolution of $\y$ is found to be accurately modeled by a symmetric Gaussian with $\sigma = 0.03$.

\begin{figure}[ht]	 
\centerline{	
\epsfysize=4.5in 	
\epsfbox{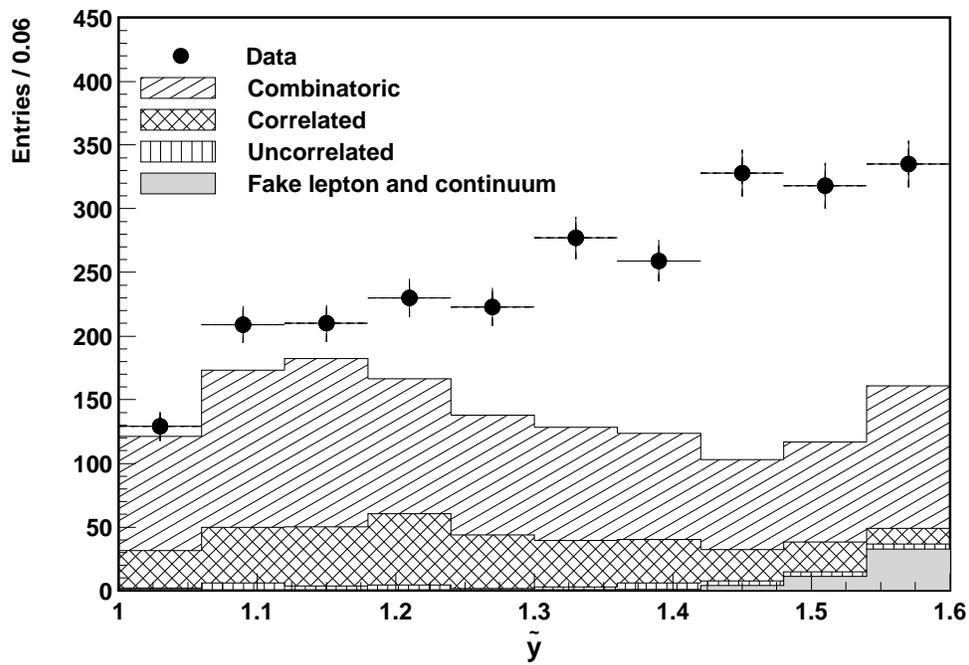}}
\caption{$\tilde{\y}$ distribution after the $\MB>5.24~{\GeVcc}$ requirement. The histogram shadings are the same as those in Figure~\ref{mbplot}.
}
\label{fg:wback}
\end{figure}

Figure~\ref{fg:wback} shows the $\tilde{\y}$ distribution for data and the estimated backgrounds.
After all backgrounds are subtracted, we perform a $\chi^2$ fit to the $\tilde{y}$ distribution to obtain $\vcbf$ and the form factor. In the fit, the $\chi^2$ function is expressed as~\cite{CLEO99}
\begin{equation}
 \chi^2=  \sum_{i=1}^{N_{bin}} \frac{(N_i^{obs} - \sum_{j=1}^{N_{bin}} \epsilon_{ij} N_j )^2}{{\sigma_{N_i}^{obs}}^2 + \sum_{j=1}^{N_{bin}} \sigma_{\epsilon_{ij}}^2 N_j^2 }, 
\end{equation}
where $N_i^{obs}$ is the yield in the $i$-th $\tilde{\y}$ bin, ${\sigma_{N_i}^{obs}}$ is the statistical error of $N_i^{obs}$, $N_{bin}$ is the number of bins and
\begin{equation}
 N_j =  N_{B\bar{B}} {\cal B} (D^+ \rt K^-\pi^+\pi^+) \tau_{\bar{B^0}} \int_{\y_{j}} d \y \frac{d \Gamma}{ d \y}
\end{equation}
is the number of decays in the $j$-th bin implied by the fit parameters. 
Here, $N_{B\bar{B}}$ is the number of $B\bar{B}$ pairs in the sample,
${\cal B} (D^+ \rt K^-\pi^+\pi^+)$ is the $D^+\rt K^-\pi^+\pi^+$ branching fraction~\cite{pdg},  $\tau_{\bar{B^0}}$ is the $\bar{B^0}$ lifetime~\cite{pdg}, and $d \Gamma / d \y$ is given by Equation 1. We assume a branching fraction ${\cal B}(\Upsilon(4S) \rt B^0\bar{B^0})=1/2$. The efficiency matrix ($\epsilon_{ij}$) accounts for the reconstruction efficiency and the smearing of ${\y}$ due to the detector resolution.  The statistical uncertainty in the efficiency matrix is represented by $\sigma_{ij}$. We use ten $\tilde{\y}$ bins over the range $1.00\le \tilde{\y} \le 1.59 $.

We use a general parametrization, 
\begin{equation}
 F_D(\y)=F_D(1)\{1-\rsqr(\y-1)+\hat{c}_D(\y-1)^2 +O(y-1)^3\}, 
\end{equation}
which is common to many analyses~\cite{CLEO99,CLEO97,ALEPH}. 
First we fit to a linear form factor neglecting terms of $O(y-1)^2$ and  higher. 
We find  $\vcbf=(3.83\pm0.35)\times 10^{-2}$ and $\rsqr = 0.69\pm0.14$ with a correlation coefficient $\rho(\vcbf,\rsqr)=0.96$, where the errors are statistical only. The $\chi^2$ is 7.6 for 8 degrees of freedom. Allowing $\hat{c}_D$ to vary in the region $\hat{c}_D>0$, we find consistent results. 

\begin{figure}[hb]	 
\centerline{	
\epsfysize=4.5in 	
\epsfbox{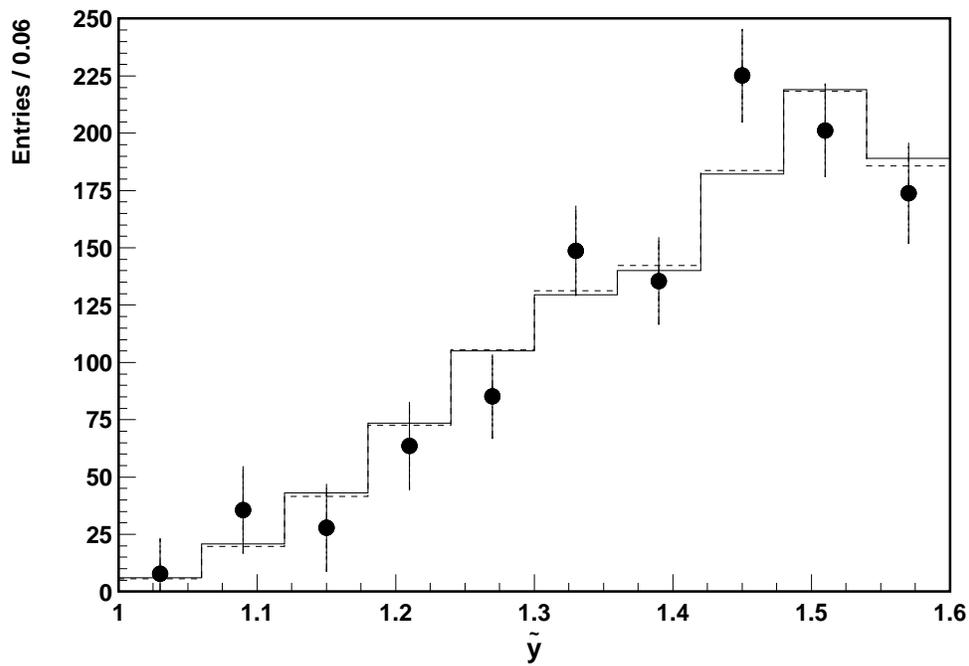}}
\caption{$\tilde{\y}$ distribution and fit. The points with error bars are data after background subtraction. The dashed histogram is the fit result for the linear form factor. The solid histogram is the fit result for the dispersion relation (Caprini $et~al.$) form factor.}
\label{fig:w_fit}
\end{figure}

\begin{figure}[hb]	 
\centerline{	
\epsfysize=4.5in 	
\epsfbox{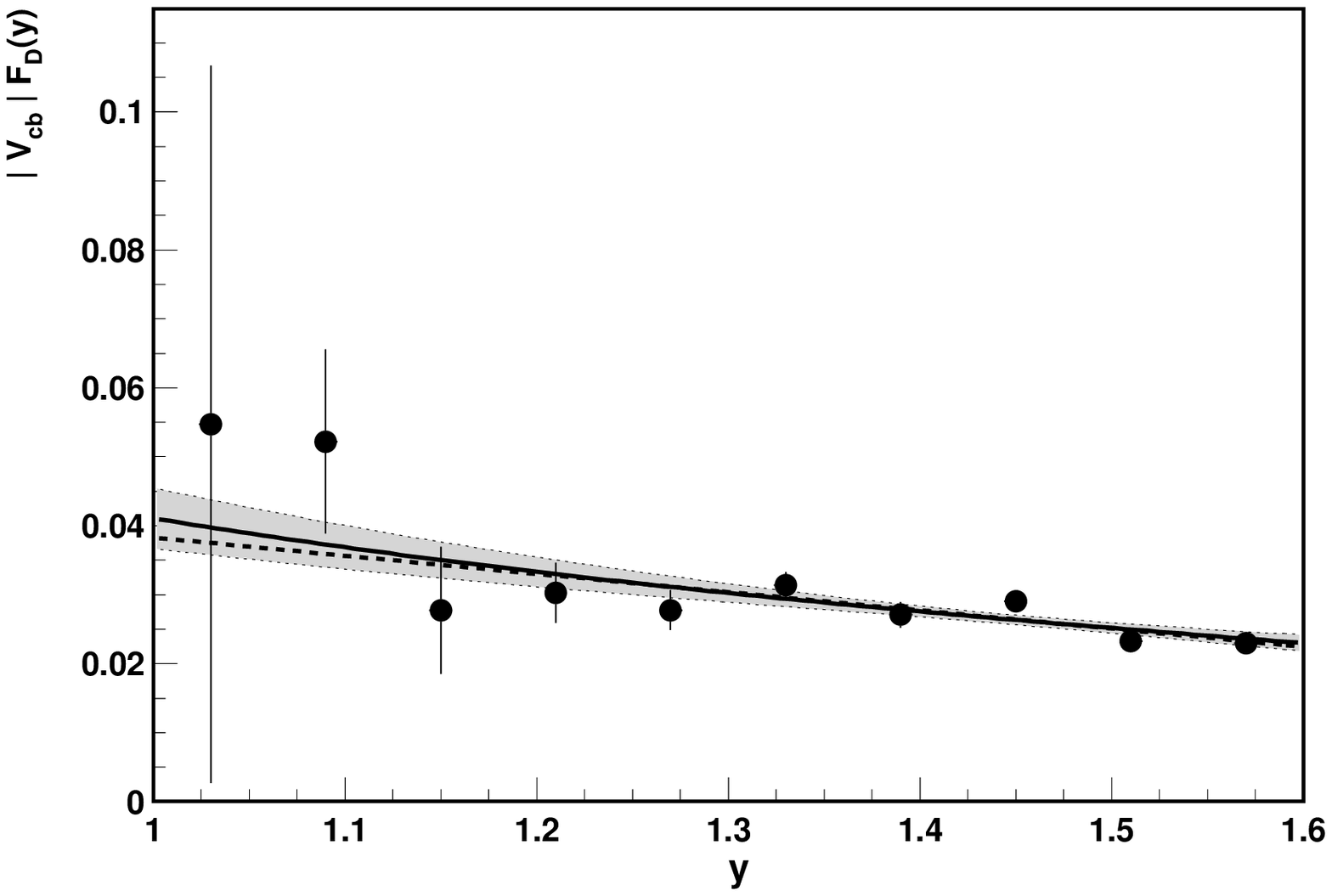}}
\caption{   $ |V_{cb}| F_D(\y) $   as a function of $\y$. The points with error bars are the data; the curves are fit results for the linear form factor (dashed) and the  Caprini $et~al.$ form factor (solid). The shaded band indicates the statistical uncertainty for the Caprini $et~al.$ form factor fit.}
\label{fig:vcbfw}
\end{figure}

Boyd $et~al.$~\cite{boyd} and Caprini $et~al.$~\cite{caprini} provide different form factor parametrizations based on QCD dispersion relations to constrain the form factor.
In each case a relation between $\rsqr$ and $\hat{c}_D$ is obtained, leaving only one free parameter, in addition to $\vcbf$, to be extracted from the fit.
They also relate higher order terms to these free parameters.
Using the form factor given by Boyd $et~al.$,
we find $\vcbf=(4.14\pm0.47) \times 10^{-2}$, $\rsqr=1.16\pm0.25$ and $\hat{c}_D=1.06\pm0.28$. 
Using the parametrization of Caprini $et~al.$, 
we find  $\vcbf=(4.11\pm0.44)\times 10^{-2}$, $\rsqr=1.12\pm0.22$ and $\hat{c}_D=1.03\pm0.23$. The fit results are shown in Figures~\ref{fig:w_fit} and~\ref{fig:vcbfw}.

We use the Caprini $et~al.$ form factor determined from
the fit to integrate the differential decay rate, $d\Gamma/d\y$, over $\y$ and
obtain the decay rate $\Gamma(\BDlnu)=(13.79\pm 0.76)\rm{~ns^{-1}}$. This decay rate leads to a branching fraction ${\cal B}(\BDlnu)=(2.13\pm0.12)\%$, where the error is statistical only. 
 The decay rate $\Gamma$ is not sensitive to the choice of form factor parametrization. 
Table~\ref{table:several_ff} summarizes the different fit results.

\section{Systematic uncertainty}
The systematic uncertainties are given in Table~\ref{SystErrDlnu}. 
The dominant uncertainty is the imperfection of the detector simulation for the neutrino reconstruction, which is determined by varying 
a number of simulation
parameters including the track finding efficiency, the track momentum resolution,  the fraction of incorrectly reconstructed tracks, the photon finding efficiency, the photon energy resolution, the charged kaon identification efficiency, the charm semileptonic decay fraction and the $K_L^0$ fraction. We take a quadratic sum of all components.

Since we rely on MC simulation to estimate the correlated backgrounds, we vary the relative fractions of $D^{*}\ell\nu$, $D^{**}\ell\nu$ and $D^{(*)}\pi\ell\nu$ within the constraints of the measured exclusive and inclusive semileptonic branching fractions~\cite{pdg}. To determine the uncertainty from the $\BDstlnu$ form factor, we vary the form factor within the uncertainties of the measurement~\cite{DstarFF} while taking into account the correlations among the parameters.

The uncorrelated, continuum
and misidentified  lepton backgrounds are very small, hence
even assuming large uncertainties in these backgrounds makes a negligible difference in our result.

The remaining contributions
to the systematic uncertainty are the uncertainty in the  $D^+$ vertexing
efficiency, and  the uncertainties in the lepton finding
efficiency, number of $\BBbar$ pairs, $D^+\to K^-\pip\pip$ branching fraction and the $\bar{B}^0$ lifetime.

We have tested the stability of the result by varying several of the
selection criteria; in each case no significant change is
observed.  Furthermore, the results obtained with the $e^-$ channel and
with the $\mu^-$ channel are consistent.

\section{Summary}
Using the missing energy and  momentum to extract kinematic information
about the undetected neutrino in the $\BDlnu$ decay, we have measured the
decay rate, $\vcbf$ and the form factor parameters for a number of
different parametrizations.  These results, with statistical errors, are
summarized in Table~\ref{table:several_ff}.  We have evaluated systematic errors of 12.5\%, 12.6\% and 18.2\% on $\vcbf$, $\rsqr$ and $\Gamma$
respectively.  Using the Caprini \emph{et al.} parametrization, we find the
decay rate of $\BDlnu$,
\[
 \Gamma (\BDlnu)=( \mygamma)~{\rm ns^{-1}},
\]
corresponding to the branching fraction,
\[
\BR (\BDlnu)= ( \mybr)\%,
\]
and we find the rate normalization,
\[
 |V_{cb}|F_{D}(1)=(\myvcbfcap)\times10^{-2}.
\]

Using  $F_{D}(1)=0.98\pm0.07$~\cite{caprini}, we find
\[
|V_{cb}| = (\myvcb)\times10^{-2}.
\]  
where the last error comes from the theoretical uncertainty of $F_D(1)$.  These  results are consistent with the result of $\bar{B}^0 \rt D^{*+} e^-\bar{\nu}$ analysis from Belle~\cite{hkjang} and with the existing measurements~ \cite{CLEO99,CLEO97,ALEPH,OPAL2000,DELPHI2001,CLEO2000dstar}

From the results of $\bar{B}^0 \rt D^{*+}e^-\bar{\nu}$ analysis at Belle,
the ratio of $F_{D}(1)$ and $F_{D^*}(1)$ and the difference between $\rsqr$ and $\hat{\rho}_{D^*}^2$ are measured to be
\[
\frac{F_D(1)}{F_{D^*}(1)} = \left\{ \begin{array}{ll} 
	                    1.12 \pm 0.12 \pm 0.12 &  \mbox{(Linear form factor)} \\
	                    1.16 \pm 0.14 \pm 0.12 &  \mbox{(Caprini {\it et~al.} form factor),}
                            \end{array}
                            \right.
\]
\[
\rsqr- \hat{\rho}_{D^*}^2  = \left\{ \begin{array}{ll} 
	                    -0.12\pm0.18\pm0.13 &  \mbox{(Linear form factor)} \\
	                    -0.23\pm0.29\pm0.20 &  \mbox{(Caprini {\it et~al.} form factor),}
                            \end{array}
                            \right.
\]

\noindent
where the first error is statistical and the second is systematic after removing the correlated error terms between the two analyses.
These results are in agreement with theoretical predictions \cite{caprini,F1V1,F1_lattice}.
                                             
\section{Acknowledgements}
We thank Benjamin Grinstein for useful discussions.  We wish to thank the KEKB accelerator group for the excellent
operation of the KEKB accelerator. We acknowledge support from the
Ministry of Education, Culture, Sports, Science, and Technology of Japan
and the Japan Society for the Promotion of Science; the Australian
Research
Council and the Australian Department of Industry, Science and
Resources; the Department of Science and Technology of India; the BK21
program of the Ministry of Education of Korea  and the Center for High Energy Physics sponsored by the KOSEF; the Polish
State Committee for Scientific Research under contract No.2P03B 17017;
the Ministry of Science and Technology of Russian Federation; the
National Science Council and the Ministry of Education of Taiwan; and
the U.S. Department of Energy.

\clearpage

\begin{table}[ht]	
\caption{\leftskip=0.25truein \rightskip=0.25truein Numbers of
signal and background events in the signal region after all event selection
requirements.}
\label{table:signal}
\begin{center}	
\begin{tabular}{c c}
\hline	
	     Total yield & $2518 \pm 50$\\\hline
	     Combinatoric background& $983 \pm 22$\\
	     Correlated background  & $349 \pm 14$\\
	     Uncorrelated background& $35 \pm 4$\\
	     Misidentified lepton background & $9 \pm 1$\\
	     Continuum background  & $43\pm7$\\\hline

	     Final signal yield & $1099 \pm 57$\\
\hline	
\end{tabular}	
\end{center}
\end{table}

\begin{table}[ht]	
  \caption[Summary of the results of the ${d\Gamma}/{d\y}$ fit]{Summary of the results of the ${d\Gamma}/{d\y}$ fit.}

  \begin{center}	
	\begin{tabular}{c c c c c c} 
\hline	

	Model  & $\vcbf / 10^{-2}$ & $\rsqr$ & $\hat{c}_D$ & $\Gamma (\rm ns^{-1})$ & $\chi^2/ \rm {dof}$ \\\hline  
	Linear & $3.83\pm0.35$&  $0.69\pm0.14$ & 0 & $13.76\pm0.76$  & 7.6/8 \\   
	Curvature&$3.83~^{+~0.46}_{-~0.35}$&  $0.69~^{+~0.42}_{-~0.15}$ & $0.00~^{+~0.59}_{-~0.00}$& $13.76\pm0.76 $ &7.6/7 \\
	Boyd $et~al.$ & $4.14\pm0.47$& $1.16\pm0.25$ & $1.06\pm0.28$& $13.78\pm0.76 $ & 8.4/8\\   
	Caprini $et~al.$ & $4.11\pm0.44$  &  $1.12\pm0.22$ & $1.03\pm0.23$& $13.79\pm 0.76$ & 8.2/8  \\   
\hline	
  \end{tabular}	
  \end{center}
 
  \label{table:several_ff}
\end{table}

\begin{table}[t]
\caption{\leftskip=0.25truein \rightskip=0.25truein Summary of
the relative systematic errors.} 
\label{SystErrDlnu}
\begin{center}
\begin{tabular}{c c c c}
\hline	
     Source of uncertainty & $\Delta \vcbf $ (\%) & $\Delta \rsqr$ (\%)& $\Delta \Gamma$ (\%)\\ \hline                        
$\nu$ reconstruction simulation&  10.6&9.7&15.5\\
Correlated background normalization &2.4  &4.4 & 1.9\\
$D^{*}$ form factor& 1.5 & 2.8& 0.9\\
Other background normalization &0.6  &1.8 & 0.4\\
$D^+$ vertexing efficiency &4.7 & 5.8 & 5.3\\
Lepton finding efficiency&1.5 & - & 3.0\\
$N_{B\bar{B}}$        & 0.5 & - & 1.0\\
$Br(D^+ \to K^-\pi^+\pi^+)$ & 3.3 & - & 6.7\\
$\tau_{\bar{B^0}}$   & 1.0 &- & 2.1\\ \hline
Total        &12.5  &12.6 &18.2 \\
\hline	
    \end{tabular}
\end{center}
\end{table}



\begin{thebibliography}{99}

\bibitem{CKM} N.~Cabibbo, Phys. Rev. Lett. {\bf 10}, 531 (1963);
M.~Kobayashi and T.~Maskawa, Prog. Theor. Phys. {\bf 49}, 652 (1973). 

\bibitem{HQET} N.~Isgur and M.~B.~Wise, Phys. Lett. {\bf B232}, 113 (1989);
M.~Neubert, Phys. Rep. {\bf 245}, 259 (1994).

\bibitem{Neubert} M.~Neubert, Phys. Lett. {\bf B264}, 455 (1991).

\bibitem{hqsymmetry} Z.~Ligeti, Y.~Nir, and M.~Neubert, Phys. Rev. {\bf D49}, 1302 (1994).


\bibitem{NIM}{
        K.~Abe {\it et al.} (Belle Collaboration),
	KEK Progress Report 2000-4 (2000),
	to be published in Nucl. Inst. and Meth. A.
}

\bibitem{kekb}{
	KEKB B Factory Design Report, KEK Report 95-7 (1995),
	unpublished; Y.~Funakoshi {\it et al.}, Proc. 2000
        European Particle Accelerator Conference, Vienna (2000).
}

\bibitem{MC} We use GEANT3 for the detector simulation: CERN Program Library Long Writeup W5013, CERN, 1993.


\bibitem{FoxWolf} G.~Fox and S.~Wolfram, Phys. Rev. Lett. $\bf{41}$,
1581 (1978).

\bibitem{CLEOPilnu} {J.~P.~Alexander {\it et al.} (CLEO Collaboration), 
Phys. Rev. Lett. {\bf 77}, 5000 (1996).}

\bibitem{DstarFF} {J.~E.~Duboscq {\it et al.} (CLEO Collaboration),
Phys. Rev. Lett. {\bf 76}, 3898 (1996).}

\bibitem{pdg}{
  	D.~E.~Groom {\it et al.} (Particle Data Group),
	Eur. Phys. J. {\bf C15}, 1 (2000).
} 

\bibitem{ISGW2} N.~Isgur and D.~Scora, Phys. Rev. {\bf D52}, 2783 (1995);
N.~Isgur, D.~Scora, B.~Grinstein, and M.~B.~Wise, Phys. Rev. {\bf
D39}, 799 (1989).  

\bibitem{goity} J.~L.~Goity and W.~Roberts, Phys. Rev. {\bf D51}, 3459 (1995).

\bibitem{BrAssumption} We assume decay rates 
${\cal B} (\bar{B} \rightarrow D_0^*\ell\bar{\nu})=0.04 \% $,
${\cal B} (\bar{B} \rightarrow D_1^*\ell\bar{\nu})=0.04 \% $,
${\cal B} (\bar{B} \rightarrow D_1\ell\bar{\nu})=0.26 \% $,
${\cal B} (\bar{B} \rightarrow D_2^*\ell\bar{\nu})=0.13 \% $,
${\cal B} (\bar{B} \rightarrow D^\prime\ell\bar{\nu})=0.01 \% $,
${\cal B} (\bar{B} \rightarrow D^{*\prime}\ell\bar{\nu})=0.09 \% $,
${\cal B} (\bar{B^0} \rightarrow D^+\pi^0\ell^-\bar{\nu})=0.32 \% $,
${\cal B} ({B}^- \rightarrow D^+\pi^-\ell^-\bar{\nu})=0.64 \% $,
${\cal B} (\bar{B} \rightarrow D^{*+}\pi^0\ell^-\bar{\nu})=0.45 \% $,
${\cal B} ({B}^- \rightarrow D^{*+}\pi^-\ell^-\bar{\nu})=0.89 \% $.



\bibitem{CLEO99} {J.~Bartelt {\it et al.} (CLEO Collaboration), 
Phys. Rev. Lett. {\bf 82} 3746 (1999).}


\bibitem{CLEO97} {M.~Athanas {\it et al.} (CLEO Collaboration),
Phys. Rev. Lett. {\bf 79}, 2208 (1997).}

\bibitem{ALEPH} {D.~Buskulic {\it et al.} (ALEPH Collaboration),
Phys. Lett. {\bf B395}, 373 (1997).} 

\bibitem{boyd} C.~G.~Boyd, B.~Grinstein and R.~F.~Lebed, Phys. Rev.{\bf~D56}, 6895, (1997).
\bibitem{caprini} I.~Caprini, L.~Lellouch and M.~Neubert, Nucl. Phys. {\bf~B530} 153, (1998).


\bibitem{hkjang} K. Abe {\it et~al.} (Belle Collaboration), 
hep-ex/0111060, submitted to Phys. Lett. {\bf~B}.

\bibitem{OPAL2000} G. Abbiendi {\it et~al.} (OPAL Collaboration), Phys. Lett.{\bf~B842}, 15, (2000).

\bibitem{DELPHI2001} P. Abreu {\it et~al.} (DELPHI Collaboration), Phys. Lett.{\bf~B510}, 55, (2001).

\bibitem{CLEO2000dstar} J. P. Alexander {\it et~al.} (CLEO Collaboration), 
hep-ex/0007052, CLEO-CONF-00-03 (2000).

\bibitem{F1V1} Z. Ligeti, Y. Nir and M.~Neubert, Phys. Rev. {\bf D49}, 1302 (1994).

\bibitem{F1_lattice} S.~Hashimoto {\it et~al.}, Phys. Rev. {\bf D61}, 14502 (1999); S.~Hashimoto {\it et~al.}, hep-ph/0110253, FERMILAB-PUB-01/317-T.
\end{thebibliography}
\end{document}